\documentclass[letterpaper]{article} 
\usepackage{aaai25}  
\usepackage{times}  
\usepackage{helvet}  
\usepackage{courier}  
\usepackage[hyphens]{url}  
\usepackage{graphicx} 
\def\UrlFont{\rm}  
\usepackage{natbib}  
\usepackage{caption} 
\usepackage{algorithm}
\usepackage{amsmath} 
\usepackage{amssymb} 
\usepackage{enumitem} 
\usepackage{algorithmic}
\usepackage{dsfont}
\usepackage{multirow}
\usepackage{booktabs}
\usepackage{array}
\usepackage{colortbl}
\usepackage{tabularx} 
\usepackage[table,xcdraw]{xcolor}
\usepackage{newfloat}
\usepackage{listings}
\DeclareCaptionStyle{ruled}{labelfont=normalfont,labelsep=colon,strut=off} 
\floatstyle{ruled}
\newfloat{listing}{tb}{lst}{}
\floatname{listing}{Listing}
\title{Adaptive Urban Planning: A Hybrid Framework for Balanced City Development}
\author{
 Pratham Singla, Ayush Singh, Adesh Gupta, Shivank Garg \\
}
\affiliations{
Vision and Language Group, Indian Institute of Technology Roorkee \\
\texttt{\{pratham\_s@me, ayush\_s@mt, adesh\_kg@ma, shivank\_g@mfs\}.iitr.ac.in } } 

\usepackage{bibentry}

\begin{document}

\maketitle

\begin{abstract}
Urban planning faces a critical challenge in balancing city-wide infrastructure needs with localized demographic preferences, particularly in rapidly developing regions. Although existing approaches typically focus on top-down optimization or bottom-up community planning, only some frameworks successfully integrate both perspectives. Our methodology employs a two-tier approach: First, a deterministic solver optimizes basic infrastructure requirements in the city region. Second, four specialized planning agents, each representing distinct sub-regions, propose demographic-specific modifications to a master planner. The master planner then evaluates and integrates these suggestions to ensure cohesive urban development. We validate our framework using a newly created dataset comprising detailed region and sub-region maps from three developing cities in India, focusing on areas undergoing rapid urbanization. The results demonstrate that this hybrid approach enables more nuanced urban development while maintaining overall city functionality. 
\end{abstract}

%

\section{Introduction}
\textit{Urban planning} is an essential field concerned with organizing city spaces to meet the needs of growing populations while balancing the demands of infrastructure, housing, transportation, and recreational facilities. Traditional top-down approaches often fail to address the diverse and localized needs of communities within a city. As cities worldwide, particularly in developing regions, experience rapid growth, there is an increasing demand for planning methods that integrate both city-wide and neighborhood-specific perspectives to create inclusive and balanced urban environments involving participation of stakeholders at multiple levels \cite{arnstein1969ladder,forester1982planning}.

India exemplifies the challenges of modern urban planning due to its demographic diversity, high population density, and history of unplanned development \cite{indianplanning}. The coexistence of historic and modern urban layouts creates issues such as congestion, inadequate green spaces, and uneven resource distribution. With the pressures of rapid urbanization, Indian cities require adaptable planning frameworks that address these complex challenges \cite{partindianplanning}.

Large Language Models (LLMs) offer a novel solution by simulating diverse stakeholder perspectives, including community members and city planners \cite{wang2024rolellmbenchmarkingelicitingenhancing}. LLMs can represent localized preferences in planning discussions, enabling strategies that balance central objectives with local demographic needs. This capability is especially relevant in India's diverse cities, where priorities differ significantly across communities.

Our work introduces a hybrid urban planning framework that leverages deterministic solvers and LLM-driven agents \cite{Wang_2024,huang2024understanding} to address the challenges of planning in rapidly urbanizing cities. Our proposed method consists of two primary components: First, the deterministic solver ensures an equitable distribution of essential infrastructure city-wide. Second, specialized LLM agents are designated to represent four distinct sub-regions, incorporating demographic-specific needs. A ``master planner” then evaluates and integrates these suggestions, ensuring the final plan aligns with city-wide goals while accommodating local preferences.

We tested our framework using data from three fast-growing cities in India. The results show that our approach handles urban infrastructure needs while considering local demographic needs. Using LLM agents can represent diverse community needs, making urban planning more inclusive and efficient.

\begin{figure*}[h]
\centering
\includegraphics[width=\linewidth, height=0.22\textheight]{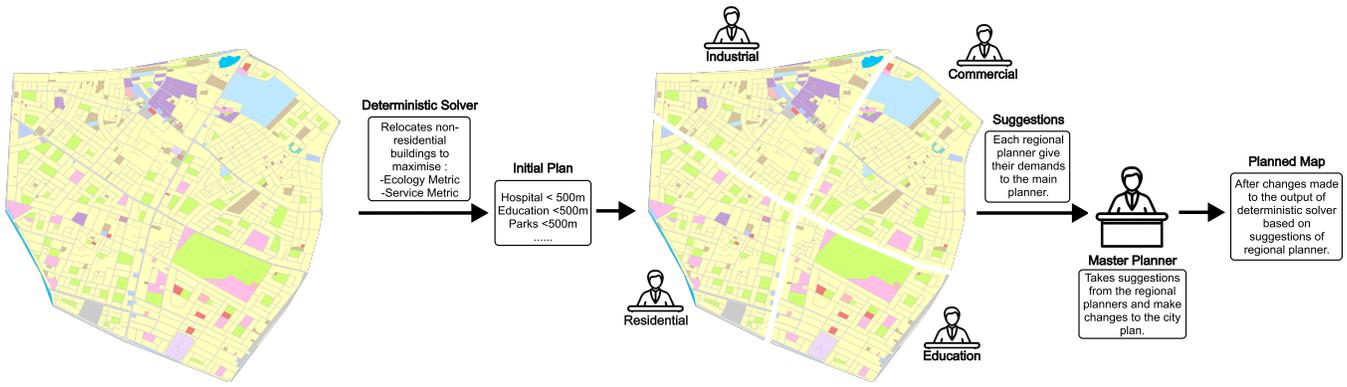} 
\caption{Workflow of the proposed urban planning framework. Integrating Deterministic Optimization, Regional Planner inputs, and Master Planner Coordination to achieve balanced and Area-Specific city layouts}
\label{approach}
\end{figure*}

\section{Related Work}

Large Language Models (LLMs) have demonstrated transformative potential in urban planning by automating complex tasks and facilitating participatory \cite{Du02042024} processes. Recent studies explore applying reasoning \cite{plaat2024reasoninglargelanguagemodels} and planning \cite{valmeekam2023planningabilitieslargelanguage} capabilities in LLMs, such as GPT-4 \cite{openai2024gpt4technicalreport}  or LLaMA3 \cite{grattafiori2024llama3herdmodels}, for specialized urban applications. For example, UrbanGPT \cite{li2024urbangptspatiotemporallargelanguage} integrates instruction-tuning and specialized decoders to enhance spatio-temporal forecasting, including traffic flow predictions. 

One significant advancement in urban planning is a participatory framework that leverages LLMs, employing role-playing agents to emulate planners and residents \cite{zhou2024large}. In this framework, LLM-based agents collaboratively design land-use plans that balance community interests with expert constraints. The system includes an LLM agent acting as the planner and numerous agents representing residents with diverse profiles and backgrounds. The planner begins by proposing an initial land-use plan. Subsequently, a simulated fishbowl discussion mechanism is employed: a subset of residents actively discusses the plan while others act as listeners. The planner then revises the plan iteratively, incorporating resident feedback to achieve a more balanced and inclusive outcome.

Building on this framework, we propose an optimized approach that employs four regional agents \cite{chen2023agentversefacilitatingmultiagentcollaboration} utilizing collaborative ability \cite{zhang2024exploringcollaborationmechanismsllm} of LLMs, each representing a specific focus area that provides suggestions to a master planner. This master planner consolidates their suggestions and revises the city plan accordingly. By reducing the number of agents, our method significantly decreases computational overhead while maintaining robust decision-making. Furthermore, our approach prioritizes meeting fundamental needs before addressing region-specific demands, ensuring a more balanced and efficient planning process than previous methods.

\section{Methodology}
The proposed methodology uses a two-tier framework for urban planning. In the first stage, a deterministic solver ensures that all residents have access to essential services and green spaces. The second stage introduces four region-specific planning agents, each advocating for the needs of their respective areas to a \textit{master planner}. This master planner evaluates these inputs and adjusts the city layout to harmonize fundamental requirements with demographic-specific needs. The overall structure of our pipeline is shown in Fig.\ref{approach}
\subsection{Deterministic Solver}
The deterministic solver leverages Genetic Algorithms (GA) \cite{forrest1996genetic,mirjalili2019genetic} to optimize urban layouts, ensuring equitable access to essential services and green spaces. The process begins with the original city plan, where essential services such as hospitals, schools, and businesses are assigned roles. To generate an initial configuration, a greedy solver creates an intermediate state by iteratively assigning elements to locations that maximize accessibility for residents. This intermediate layout provides a near-optimal starting point for the GA to refine further.


The GA uses two main steps: \textit{mutation} and \textit{selection}. Mutation creates new layouts by randomly swapping roles between locations, helping to explore different options and avoid premature convergence. Selection picks the best layouts using a tournament method, this favors high-performing layouts while maintaining diversity to explore alternative layouts.

Each layout's fitness is evaluated using two metrics: 
\begin{enumerate}
\item \textbf{Service Accessibility Metric:} Measures the availability of essential services within 500 meters of residential areas.
\item \textbf{Ecological Proximity Metric:} Evaluates access to green spaces within 300 meters, reflecting urban livability and resident well-being.
\end{enumerate}

The GA iteratively improves the layout across successive generations by ranking layouts based on fitness, retaining the top-performing configurations, and generating new ones through mutation. This process continues until fitness improvements plateau or a predefined number of generations is reached, indicating convergence.

The outputs of the deterministic solver include an optimized urban layout that maximizes accessibility to essential services and green spaces. Detailed formulation is in Appendix.

\subsection{Regional Adaptation via Dual-Planners}
Building on the deterministic solver's output, we employ a dual-planner approach to address regional and demographic needs. The region is divided into four sub-regions, each guided by a regional planner responsible for advocating for its area-specific requirements. The master planner then reviews these proposals, integrating them into the city layout to balance local priorities with city-wide objectives. This framework ensures that regional needs are addressed without compromising structural integrity or urban efficiency.

\subsubsection{Master Planner}:
It operates with a city-wide perspective, maximizing accessibility and achieving a balanced distribution of facilities. It prevents clustering in central areas and avoids over-dispersing facilities toward city edges, ensuring even coverage across the urban area.

Adhering to a minimal-change policy, the master planner makes essential layout adjustments only when necessary. These include reassigning vacant land for high-priority facilities, adding essential services in underserved areas, or swapping facility types to maintain efficient resource distribution. This strategy preserves the city’s structural integrity while meeting overarching urban planning goals.

\subsubsection{Regional Planners}: It complements the master planner by addressing the sub-region's specific demographic and functional needs. Each regional planner is designated to focus on one of four demographic roles: Industrial, Educational, Commercial, and Residential, chosen for their relevance to urban functionality. 
\begin{itemize}
\item \textbf{Industrial zones} prioritize factories, warehouses, and logistics hubs.
\item \textbf{Educational zones} emphasize schools, universities, and student housing.
\item \textbf{Commercial zones} focus on offices, retail spaces, and business-support infrastructure.
\item \textbf{Residential zones} ensure access to housing, community spaces, and daily amenities.
\end{itemize}

Tailoring facilities to these roles ensures that every part of the city is equipped to meet the unique needs of its population, fostering a well-integrated and sustainable urban environment.

We used GPT4o-Mini\footnote{https://openai.com/index/gpt-4o-mini-advancing-cost-efficient-intelligence/} for both master and regional planners, as it offers adaptability for handling complex urban planning tasks. Together, the \textit{master} and \textit{regional} planners foster a coordinated city plan that upholds structural integrity and ecological balance while adapting to the particular needs of different areas.

\section{Dataset}
Our dataset comprises thematic maps, essential for precise analysis of existing urban layouts. These maps, sourced from Bhuvan AMRUT 4K \cite{BhuvanAmrut} web services, distinctly mark urban areas with categories like residential, government property, commercial zones, transportation, green spaces, educational institutions, and more. Such differentiation aids our framework by providing a clear view of diverse land uses within city boundaries.

From the 238 available AMRUT city maps, we selected Kanpur, Lucknow, and Raipur for evaluation. These cities were chosen for their varied urban contexts, including industrial zones, administrative hubs, and rapidly growing urban areas. This diversity ensures that our framework is tested across different urban planning challenges, highlighting its adaptability and effectiveness.

We extracted map data via Bhuvan’s API by manually setting coordinates for each target area. Using connected component analysis with optimal parameters, we isolated regions within the maps based on land use type. This granular extraction process provides a high-resolution foundation for applying our integrated planning model across distinct urban contexts. For extraction details, refer to Appendix.

\section{Evaluation}

To assess the effectiveness of our proposed framework, we use three key metrics: \textit{Service Accessibility}, \textit{Ecological Coverage}, and \textit{Resident Satisfaction}. Together, these metrics assess the accessibility of public services, the availability of green spaces, and the fulfillment of residents' demographic-specific needs, emphasizing the framework’s ability to create accessible, ecologically balanced, and resident-centric urban environments.

\begin{table*}[htbp]
\centering
\begin{tabularx}{\textwidth}{l|XXX|XXX|XXX}
\toprule
\textbf{Metrics} & 
\multicolumn{3}{c|}{\textbf{Kanpur}} & 
\multicolumn{3}{c|}{\textbf{Lucknow}} & 
\multicolumn{3}{c}{\textbf{Raipur}} \\
\cmidrule{2-10}
 & \textbf{Stage1} & \textbf{Stage2} & \textbf{Stage3} & 
\textbf{Stage1} & \textbf{Stage2} & \textbf{Stage3} & 
\textbf{Stage1} & \textbf{Stage2} & \textbf{Stage3} \\
\midrule
\textbf{Service} & 0.791 & 0.892 & \textbf{0.916} & 0.855 & 0.908 & \textbf{0.943} & 0.783 & 0.922 & \textbf{0.948} \\
\textbf{Ecology} & 0.868 & 0.899 & \textbf{0.899} & 0.709 & 0.946 & \textbf{0.946} & 0.825 & 0.842 & \textbf{0.842} \\
\textbf{Satisfaction} & 0.307 & 0.327 & \textbf{0.489} & 0.294 & 0.326 & \textbf{0.683} & 0.372 & 0.377 & \textbf{0.615} \\
\bottomrule
\end{tabularx}
\caption{Performance metrics across cities and planning stages (Stage 1: Baseline, Stage 2: Optimized Layout by Deterministic solver, Stage 3: Final Integration)}
\label{research_results}
\end{table*}

\subsection{Service Accessibility}

The \textit{Service Accessibility} metric evaluates how efficiently essential services are distributed within residential areas. It measures the proportion of essential services (e.g., education, healthcare, workplaces, shopping, and recreation) accessible within a 500-meter radius of resident's homes, with values ranging from 0 to 1, where higher values represent better service accessibility.

The metric is computed as follows:
\begin{enumerate}
    \item For each resident \(m\), the minimum distance \(d(m, j)\) to access a facility of type \(j\) is determined:
    \begin{equation}
    d(m, j) = \underset{1 \leq i \leq k_j}{\min} \textit{EucDis}(L_m, P_{j, i})
    \end{equation}
    where \(L_m\) is the resident’s location, and \(P_{j, k}\) denotes the \(k\)-th facility of type \(j\).

    \item The overall \textit{Service Accessibility} metric aggregates these values for all residents \(n_m\) and service types \(n_j\):
    \begin{equation}
        \text{Service} = \frac{1}{n_m} \sum_{m=1}^{n_m} \frac{1}{n_j} \sum_{j=1}^{n_j} \mathds{1}[d(m, j) < 500],
    \end{equation}
    where \(\mathds{1}[d(m, j) < 500]\) is an indicator function, returning 1 if the distance is less than 500 meters, and 0 otherwise.
\end{enumerate}

\subsection{Ecological Coverage}

The \textit{Ecological Coverage} metric measures the availability of parks and green spaces, which play a critical role in promoting the health and well-being of urban residents. This metric evaluates the proportion of residents who live within a 300-meter radius of parks or open spaces, aligning with global standards for urban green accessibility.

The metric calculation is as follows:
\begin{enumerate}
    \item The \textit{Ecological Service Area (ESA)} is defined as the combined buffer zones extending 300 meters around each park or green space:
    \begin{equation}
    \text{ESR} = \bigcup_{i=1}^{k} \text{Buffer}(P_{\text{park}, i}, 300),
\end{equation}

    where \(P_{\text{park}, k}\) represents the \(k\)-th park or green space.

    \item The \textit{Ecological Coverage} metric is computed as the proportion of residents \(L_m\) located within the ESA:
    \begin{equation}
        \text{Ecological Coverage} = \frac{1}{n_m} \sum_{m=1}^{n_m} \mathds{1}[L_m \in \text{ESA}],
    \end{equation}
    where \(\mathds{1}[L_m \in \text{ESA}]\) returns 1 if the resident is within the buffer zone and 0 otherwise.
\end{enumerate}

\subsection{Satisfaction}

The \textit{Satisfaction} metric evaluates how effectively the urban layout fulfills the specific needs of residents in different demographic sub-regions. Unlike the previous two metrics, this metric considers the unique requirements of each sub-region, such as educational facilities in academic zones or healthcare in posh neighborhoods, ensuring a more customized urban planning approach. This metric ranges from 0 to 1, with higher values indicating better alignment between urban layouts and resident’s specific needs.

\begin{enumerate}
    \item Each resident \(m\) in a sub-region is assigned a set of prioritized needs \(J_m\), representing 3-5 most critical land-use categories for that demographic goal. The satisfaction level for an individual resident \(m\) is calculated as:
    \begin{equation}
        S_m = \frac{1}{n_j} \sum_{j \in J_m} \mathds{1}[d(m, j) < 800],
    \end{equation}
    where \(d(m, j)\) is the minimum distance from the resident to a facility of type \(j\), and \(\mathds{1}[d(m, j) < 800]\) indicates whether this distance is within 800 meters.

    \item The overall \textit{Satisfaction Metric} is then computed by aggregating the satisfaction values across all residents \(n_m\) in the region:
    \begin{equation}
        \text{Satisfaction} = \frac{1}{n_m} \sum_{m=1}^{n_m} S_m
    \end{equation}
\end{enumerate}

Together, the \textit{Service Accessibility}, \textit{Ecological Coverage}, and \textit{Satisfaction} metrics evaluate urban layouts by balancing accessibility, environmental sensitivity, and demographic inclusivity. These metrics demonstrate our framework’s alignment with the concept of a ``15-minute city" \cite{smartcities4010006}, ensuring essential services and green spaces are within walking or cycling distance, fostering sustainable and resident-focused urban spaces for rapidly urbanizing regions.

\section{Results}
Table \ref{research_results} illustrates progressive improvements across the three key metrics over the planning stages. Stage 1 established the baseline, revealing disparities in accessibility and sustainability and emphasizing the need for integrated planning. Applying the deterministic solver in Stage 2 led to significant gains in both Service Accessibility and Ecological Coverage. This step ensured a more balanced distribution of essential services and green spaces, laying a solid foundation for livable urban environments. The final Stage 3, which incorporated inputs from specialized regional planning agents and coordination by the master planner, further enhanced all metrics. This stage addressed localized demographic needs while maintaining city-wide balance, substantially improving resident satisfaction. Additional regional results are in Appendix.

These results demonstrate that our hybrid methodology, which integrates systematic optimization with regional customization, effectively supports the development of accessible, ecologically sustainable, and community-oriented urban spaces. This approach is particularly beneficial in rapidly developing cities, where balancing diverse needs with overall urban functionality is critical.

\section{Conclusion}
This work presents a hybrid urban planning framework that optimizes city-wide infrastructure with localized demographic needs. By employing a two-tier methodology consisting of a deterministic solver and region-specific planning agents, our approach balances functional efficiency and community-specific requirements.
The evaluation using data from three rapidly urbanizing Indian cities demonstrates notable improvements in Service Accessibility, Ecological Coverage, and Resident Satisfaction across successive planning stages. The results highlight the advantages of combining systematic optimization with adaptive regional planning to create sustainable, inclusive, and livable urban environments.

Our framework offers a scalable and practical solution for addressing the challenges of urbanization in developing regions, ensuring that urban growth aligns with both ecological priorities and diverse community needs.

\bibliography{aaai25}

\begin{thebibliography}{20}
\providecommand{\natexlab}[1]{#1}

\bibitem[{Arnstein(1969)}]{arnstein1969ladder}
Arnstein, S.~R. 1969.
\newblock A Ladder Of Citizen Participation.
\newblock \emph{Journal of the American Institute of Planners}, 35(4): 216--224.

\bibitem[{Bhuvan(2022)}]{BhuvanAmrut}
Bhuvan. 2022.
\newblock AMRUT.

\bibitem[{Chen et~al.(2023)Chen, Su, Zuo, Yang, Yuan, Chan, Yu, Lu, Hung, Qian, Qin, Cong, Xie, Liu, Sun, and Zhou}]{chen2023agentversefacilitatingmultiagentcollaboration}
Chen, W.; Su, Y.; Zuo, J.; Yang, C.; Yuan, C.; Chan, C.-M.; Yu, H.; Lu, Y.; Hung, Y.-H.; Qian, C.; Qin, Y.; Cong, X.; Xie, R.; Liu, Z.; Sun, M.; and Zhou, J. 2023.
\newblock AgentVerse: Facilitating Multi-Agent Collaboration and Exploring Emergent Behaviors.
\newblock arXiv:2308.10848.

\bibitem[{Du et~al.(2024)Du, Ye, Jankowski, Sanchez, and Mai}]{Du02042024}
Du, J.; Ye, X.; Jankowski, P.; Sanchez, T.~W.; and Mai, G. 2024.
\newblock Artificial intelligence enabled participatory planning: a review.
\newblock \emph{International Journal of Urban Sciences}, 28(2): 183--210.

\bibitem[{Forester(1982)}]{forester1982planning}
Forester, J. 1982.
\newblock Planning in the Face of Power.
\newblock \emph{Journal of the American Planning Association}, 48(1): 67--80.

\bibitem[{Forrest(1996)}]{forrest1996genetic}
Forrest, S. 1996.
\newblock Genetic algorithms.
\newblock \emph{ACM computing surveys (CSUR)}, 28(1): 77--80.

\bibitem[{Grattafiori et~al.(2024)Grattafiori, Dubey, Jauhri, Pandey, Kadian, Al-Dahle, Letman, and Mathur}]{grattafiori2024llama3herdmodels}
Grattafiori, A.; Dubey, A.; Jauhri, A.; Pandey, A.; Kadian, A.; Al-Dahle, A.; Letman, A.; and Mathur, A. 2024.
\newblock The Llama 3 Herd of Models.
\newblock arXiv:2407.21783.

\bibitem[{Huang et~al.(2024)Huang, Liu, Chen, Wang, Wang, Lian, Wang, Tang, and Chen}]{huang2024understanding}
Huang, X.; Liu, W.; Chen, X.; Wang, X.; Wang, H.; Lian, D.; Wang, Y.; Tang, R.; and Chen, E. 2024.
\newblock Understanding the Planning of LLM Agents: A Survey.
\newblock \emph{arXiv preprint}, arXiv:2402.02716.

\bibitem[{Kumar and Prakash(2016)}]{partindianplanning}
Kumar, A.; and Prakash, P. 2016.
\newblock Public participation in planning in India.

\bibitem[{Li et~al.(2024)Li, Xia, Tang, Xu, Shi, Xia, Yin, and Huang}]{li2024urbangptspatiotemporallargelanguage}
Li, Z.; Xia, L.; Tang, J.; Xu, Y.; Shi, L.; Xia, L.; Yin, D.; and Huang, C. 2024.
\newblock UrbanGPT: Spatio-Temporal Large Language Models.
\newblock arXiv:2403.00813.

\bibitem[{Mirjalili and Mirjalili(2019)}]{mirjalili2019genetic}
Mirjalili, S.; and Mirjalili, S. 2019.
\newblock Genetic algorithm.
\newblock \emph{Evolutionary algorithms and neural networks: theory and applications}, 43--55.

\bibitem[{Moreno et~al.(2021)Moreno, Allam, Chabaud, Gall, and Pratlong}]{smartcities4010006}
Moreno, C.; Allam, Z.; Chabaud, D.; Gall, C.; and Pratlong, F. 2021.
\newblock Introducing the “15-Minute City”: Sustainability, Resilience and Place Identity in Future Post-Pandemic Cities.
\newblock \emph{Smart Cities}, 4(1): 93--111.

\bibitem[{OpenAI et~al.(2024)OpenAI, Achiam, Adler, Agarwal, Ahmad, Akkaya, and Aleman}]{openai2024gpt4technicalreport}
OpenAI; Achiam, J.; Adler, S.; Agarwal, S.; Ahmad, L.; Akkaya, I.; and Aleman, F.~L. 2024.
\newblock GPT-4 Technical Report.
\newblock arXiv:2303.08774.

\bibitem[{Plaat et~al.(2024)Plaat, Wong, Verberne, Broekens, van Stein, and Back}]{plaat2024reasoninglargelanguagemodels}
Plaat, A.; Wong, A.; Verberne, S.; Broekens, J.; van Stein, N.; and Back, T. 2024.
\newblock Reasoning with Large Language Models, a Survey.
\newblock arXiv:2407.11511.

\bibitem[{Ranjan(2023)}]{indianplanning}
Ranjan, N. 2023.
\newblock Economic Planning in Practice: Indian Experience and NITI Aayog.

\bibitem[{Valmeekam et~al.(2023)Valmeekam, Marquez, Sreedharan, and Kambhampati}]{valmeekam2023planningabilitieslargelanguage}
Valmeekam, K.; Marquez, M.; Sreedharan, S.; and Kambhampati, S. 2023.
\newblock On the Planning Abilities of Large Language Models : A Critical Investigation.
\newblock arXiv:2305.15771.

\bibitem[{Wang et~al.(2024{\natexlab{a}})Wang, Ma, Feng, Zhang, Yang, Zhang, Chen, Tang, Chen, Lin, Zhao, Wei, and Wen}]{Wang_2024}
Wang, L.; Ma, C.; Feng, X.; Zhang, Z.; Yang, H.; Zhang, J.; Chen, Z.; Tang, J.; Chen, X.; Lin, Y.; Zhao, W.~X.; Wei, Z.; and Wen, J. 2024{\natexlab{a}}.
\newblock A survey on large language model based autonomous agents.
\newblock \emph{Frontiers of Computer Science}, 18(6).

\bibitem[{Wang et~al.(2024{\natexlab{b}})Wang, Peng, Que, Liu, Zhou, Wu, Guo, Gan, Ni, Yang, Zhang, Zhang, Ouyang, Xu, Huang, Fu, and Peng}]{wang2024rolellmbenchmarkingelicitingenhancing}
Wang, Z.~M.; Peng, Z.; Que, H.; Liu, J.; Zhou, W.; Wu, Y.; Guo, H.; Gan, R.; Ni, Z.; Yang, J.; Zhang, M.; Zhang, Z.; Ouyang, W.; Xu, K.; Huang, S.~W.; Fu, J.; and Peng, J. 2024{\natexlab{b}}.
\newblock RoleLLM: Benchmarking, Eliciting, and Enhancing Role-Playing Abilities of Large Language Models.
\newblock arXiv:2310.00746.

\bibitem[{Zhang et~al.(2024)Zhang, Xu, Zhang, Liu, Hooi, and Deng}]{zhang2024exploringcollaborationmechanismsllm}
Zhang, J.; Xu, X.; Zhang, N.; Liu, R.; Hooi, B.; and Deng, S. 2024.
\newblock Exploring Collaboration Mechanisms for LLM Agents: A Social Psychology View.
\newblock arXiv:2310.02124.

\bibitem[{Zhou et~al.(2024)Zhou, Lin, Jin, and Li}]{zhou2024large}
Zhou, Z.; Lin, Y.; Jin, D.; and Li, Y. 2024.
\newblock Large language model for participatory urban planning.
\newblock \emph{arXiv preprint arXiv:2402.17161}.

\end{thebibliography}

\clearpage
\appendix
\section{Formulation of Deterministic Solver}
\label{deterministic_solver}
\begin{itemize}
    \item \( S_0 \) : Initial game state (mapping of regions to roles, initially set to ``None").
    \item \( P \) : Set of players, representing the non-residential types to be assigned to regions.
    \item \( C \) : Centroids dictionary (coordinates of the center of each region).
    \item \( L \) : Move limits dictionary, where \( L[p] \) denotes the number of assignments allowed for player \( p \).
    \item \( S_{\text{final}} \) : Final optimized game state after the genetic algorithm process.
    \item \( r^* \) : Region selected for assignment based on the highest return value in the greedy phase.
    \item \( \mathcal{P} \) : Population of layout configurations in the genetic algorithm.
    \item \( N \) : Population size for the genetic algorithm.
    \item \( G \) : Number of generations in the genetic algorithm.
    \item \( k \) : Number of top layouts selected for the next generation.
    \item \( S^* \) : Layout with the highest fitness value after the genetic algorithm optimization.
    \item \( \texttt{calculate\_return} \) : Function used to calculate the return value for assigning a region to a player based on service and ecology metrics.
    \item \( \texttt{fitness\_function} \) : Function that evaluates the fitness of a layout based on service accessibility and ecological proximity.
    \item \( \texttt{mutate} \) : Function that applies random swaps to create new variations of a layout.
    \item \( \texttt{initialize\_population} \) : Function that generates the initial population for the genetic algorithm using random swaps.
\end{itemize}
\begin{enumerate}
    
    \item \textbf{Input:}
    \begin{itemize}
        \item Initial game state $S_0$ (mapping of regions to roles).
        \item Players $P$ (list of non-residential types to assign).
        \item Centroids dictionary $C$ (coordinates of region centers).
        \item Move limits $L$ (number of assignments allowed for each player).
    \end{itemize}
    \item \textbf{Phase 1: Greedy Assignment.}
    \begin{enumerate}
        \item Initialize the game state $S \gets S_0$.
        \item While unassigned regions remain and any $L[p] > 0$ for $p \in P$:
        \begin{enumerate}
            \item For each player $p \in P$:
            \begin{enumerate}
                \item If $L[p] = 0$, continue to the next player.
                \item Find the region $r^*$ that maximizes the return value using \texttt{calculate\_return}.
                \item Assign $r^*$ to $p$: $S[r^*] \gets p$.
                \item Decrease the move limit: $L[p] \gets L[p] - 1$.
            \end{enumerate}
        \end{enumerate}
        \item Output intermediate layout $S_{\text{greedy}}$.
    \end{enumerate}
    \item \textbf{Phase 2: Genetic Algorithm Optimization.}
    \begin{enumerate}
        \item Initialize population $\mathcal{P}$ of size $N$ using $S_{\text{greedy}}$ and random swaps.
        \item For each generation $g \in \{1, 2, \ldots, G\}$:
        \begin{enumerate}
            \item Evaluate the fitness of each layout $S \in \mathcal{P}$ using \texttt{fitness\_function}.
            \item Select the top $k$ layouts to carry forward.
            \item Mutate layouts to create $N - k$ new layouts and add to the next generation.
        \end{enumerate}
        \item Output the layout $S^*$ with the highest fitness in $\mathcal{P}$.
    \end{enumerate}
    \item \textbf{Output:} $S_{\text{final}} \gets S^*$, the layout maximizing service accessibility and ecological proximity.
\end{enumerate}

\section{Extraction of infomation from image}
\begin{figure}[h]
\centering
\includegraphics[width=\linewidth, height=0.18\textheight]{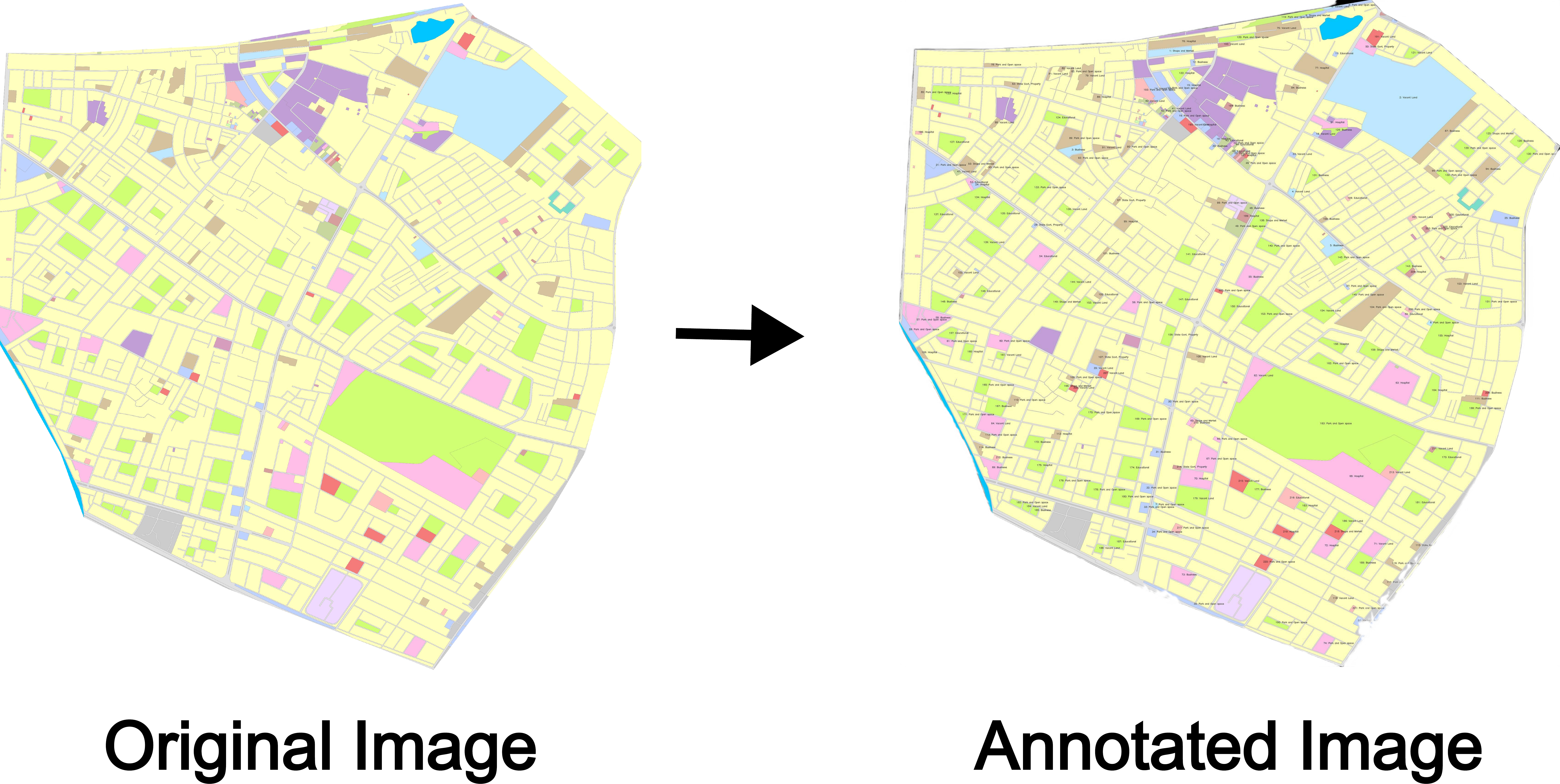}
\caption{Conversion of the original map image into the annotated map image showcasing land-use categorization through color-based segmentation}
\label{annotation}
\end{figure}

We used a predefined color legend (Table \ref{legend}) to extract information from the image to categorize various land regions on a geographic map. Each land-use type, such as Residential, Business, Educational, and others, was associated with a specific color, enabling efficient map segmentation based on these color codes. The map, as illustrated in Figure \ref{annotation}, was first converted into the HSV (Hue, Saturation, Value) color space, facilitating easier color segmentation by defining precise color ranges for each land type. This transformation allowed for identifying pixels corresponding to specific regions, effectively distinguishing different land-use categories.

\begin{table}[h!]
\centering
\begin{tabular}{|c|c|p{3.5cm}|} 
\hline
\textbf{Colour} & \textbf{Type} & \textbf{Associated Structures} \\
\hline
\cellcolor[RGB]{255,255,190}  & Residential & Houses, Apartments, Villas \\
\hline
\cellcolor[RGB]{194,231,252}  & State Govt. Property & Government offices, Emergency services (e.g., police, fire stations) \\
\hline
\cellcolor[RGB]{192,209,254}  & Business & Commercial buildings, Office spaces, Retail stores \\
\hline
\cellcolor[RGB]{255,235,190}  & Public Utilities & Water treatment plants, Sewage systems, Electricity stations \\
\hline
\cellcolor[RGB]{200,214,157}  & Shops and Market & Markets, Grocery stores, Shopping malls \\
\hline
\cellcolor[RGB]{254,191,229}  & Educational & Schools, Universities, Libraries, Educational centers \\
\hline
\cellcolor[RGB]{214,194,158}  & Vacant Land & Open fields, Unused land \\
\hline
\cellcolor[RGB]{210,255,116}  & Park and Open Space & Public parks, Playgrounds, Green spaces \\
\hline
\cellcolor[RGB]{255,190,190}  & Hospital & Hospitals, Clinics, Healthcare facilities \\
\hline
\end{tabular}
\caption{Pre-defined color legend for categorizing land-use types, associating each color with specific structures to support map segmentation and spatial analysis}
\label{legend}
\end{table}

\begin{table*}[htbp]
\centering
\begin{tabularx}{\textwidth}{l|XXX|XXX|XXX}
\toprule
\textbf{Metrics} &
\multicolumn{3}{c|}{\textbf{Kanpur}} &
\multicolumn{3}{c|}{\textbf{Lucknow}} &
\multicolumn{3}{c}{\textbf{Raipur}} \\
\cmidrule{2-10}
& \textbf{Stage1} & \textbf{Stage2} & \textbf{Stage3} &
\textbf{Stage1} & \textbf{Stage2} & \textbf{Stage3} &
\textbf{Stage1} & \textbf{Stage2} & \textbf{Stage3} \\
\midrule
\textbf{Service} & 0.432 & 0.644 & \textbf{0.710} & 0.749 & 0.860 & \textbf{0.895} & 0.812 & 0.859 & \textbf{0.926} \\
\textbf{Ecology} & 0.840 & 0.951 & \textbf{0.951} & 0.627 & 0.656 & \textbf{0.656} & 0.485 & 0.617 & \textbf{0.617} \\
\textbf{Satisfaction} & 0.325 & 0.355 & \textbf{0.507} & 0.294 & 0.439 & \textbf{0.765} & 0.510 & 0.495 & \textbf{0.653} \\
\bottomrule
\end{tabularx}
\caption{Performance metrics across additional regions of Kanpur, Lucknow and Raipur, demonstrating the generalizability of the proposed hybrid planning framework (Stage 1: Baseline, Stage 2: Optimized Layout by Deterministic solver, Stage 3: Final Integration)}
\label{extra_eval}
\end{table*}

We then performed connected component analysis on the map to identify distinct regions. Each identified region was labeled, and its area and centroid were calculated. Only regions with a minimum area threshold were considered for further analysis. For each valid region, a mask was generated, and relevant details, including the land type, label, area, and centroid, were recorded. The data was then organized into a structured format, making it suitable for further urban planning, environmental assessment, or other relevant applications. This process enabled the efficient extraction and categorization of land-use regions from the map, supporting various spatial analysis tasks.

\section{Sub-Region Extraction}
\label{sub_region}
\begin{figure}[h]
\centering
\includegraphics[width=\linewidth, height=0.35\textheight]{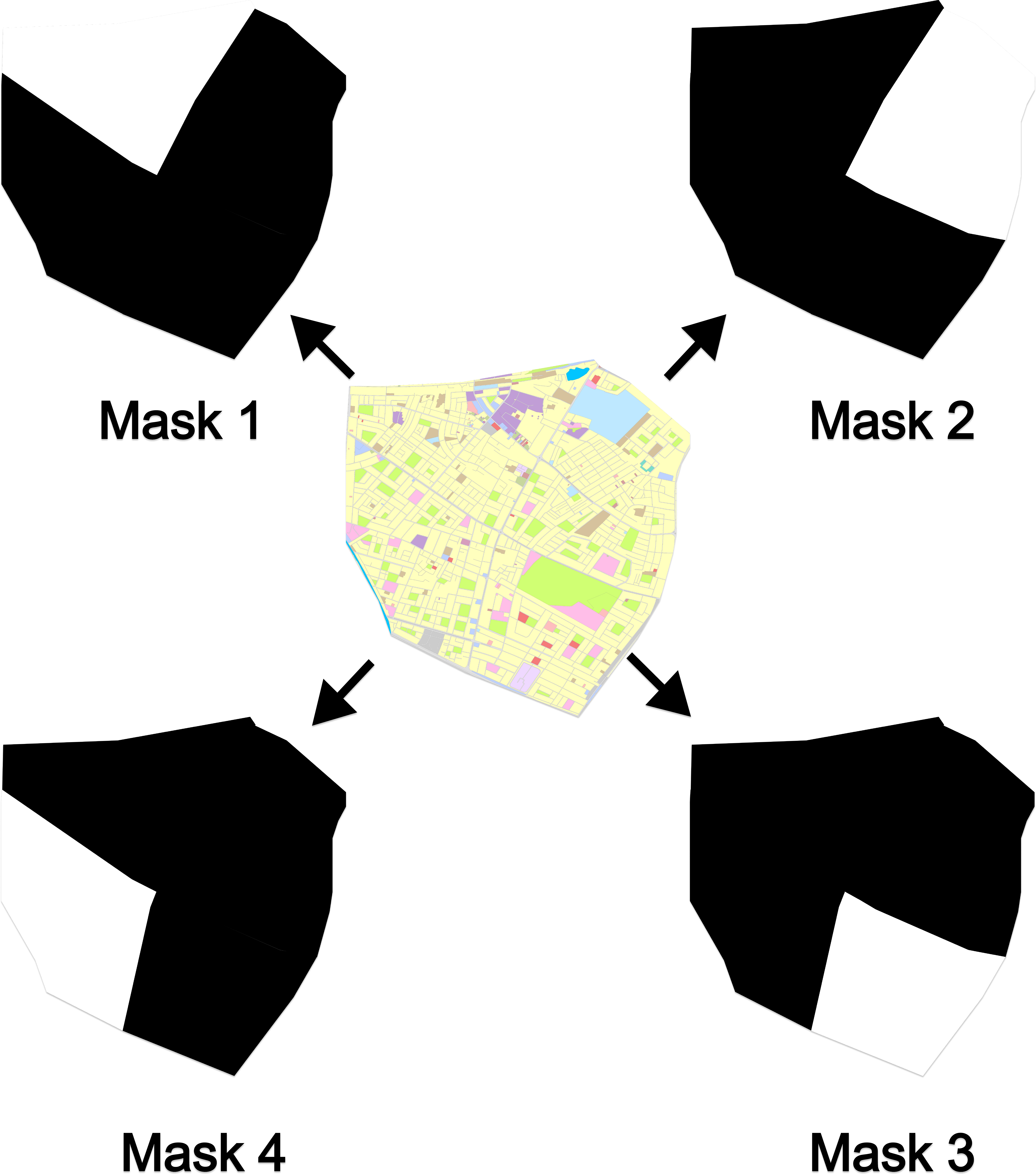}
\caption{Utilization of mask images to validate centroids of land-use regions, showing the original map and corresponding masks defining valid sub-regions}
\label{sub-region_mask}
\end{figure}

A mask image representing predefined regions on a map is utilized to filter and validate centroids of land-use regions to extract sub-regions. The mask image, as shown in Figure \ref{sub-region_mask}, is loaded in grayscale, where white areas correspond to valid regions of interest. The dimensions of the mask image are verified to ensure proper alignment with the spatial data. A function is then defined to check if a given region's centroid falls within the mask's white area. This is done by converting the centroid's coordinates to integers and checking if they lie within the image boundaries and if the pixel at that location is white (indicating a valid region).

The filtering process is applied to the centroids of all regions in the dataset, and only those regions whose centroids fall within the white area of the mask are retained. This ensures that only relevant regions located within predefined valid areas are considered for further analysis or processing. The result is a refined dataset containing only the regions that meet the criteria, enabling more focused and accurate urban or environmental assessments.

\section{Further Evaluation}
\label{extra_eval}
To further evaluate the robustness and generalizability of our hybrid urban planning framework, we applied the approach to additional regions within the cities of Kanpur, Lucknow, and Raipur. The performance metrics for these extended evaluations, summarized in Table \ref{extra_eval}, demonstrate consistent improvements across all three key metrics—Service Accessibility, Ecological Coverage, and Resident Satisfaction across the three planning stages.
These results indicate the framework’s ability to generalize across different urban environments, demonstrating its effectiveness in addressing the complexities of urban planning in diverse regions. By combining systematic optimization with community-centric customization, the proposed methodology establishes a scalable and practical model for sustainable urban development.

\end{document}